\documentclass[prd,superscriptaddress,a4paper,showpacs,showkeys,10pt,nofootinbib]{revtex4}
\usepackage{graphicx}
\usepackage{graphics}
\usepackage{epsfig}
\usepackage{dcolumn}
\usepackage{bm}
\usepackage{multirow}
\usepackage{tabularx}
\usepackage{hyperref}
\usepackage{commath}
\usepackage{epstopdf}
\usepackage[T1]{fontenc}
\usepackage{geometry}
\usepackage{color}

\providecommand{\superchic}{{\sc superchic}}

\newcommand{\sqrtsnn}{\sqrt{s_{_{\textsc{nn}}}}}

\def\be{\begin{equation}}
\def\ee{\end{equation}}

\providecommand{\ee}{e$^+$e$^-$}

\providecommand{\tabularnewline}{\\}

\makeatother

\begin{document}
%
%
\title{Searching for axionlike particles with low masses in $pPb$ and $PbPb$ collisions}


\author{V. P. Gon\c calves}

\email[]{barros@ufpel.edu.br}

\affiliation{Instituto de F\'{\i}sica e Matem\'atica, Universidade Federal de
Pelotas (UFPel),\\
Caixa Postal 354, CEP 96010-090, Pelotas, RS, Brazil}

\author{D. E. Martins}

\email[]{dan.ernani@gmail.com}

\affiliation{No affiliation}

\author{M. S. Rangel}

\email[]{rangel@if.ufrj.br}

\affiliation{Instituto de F\'isica, Universidade Federal do Rio de Janeiro (UFRJ), 
Caixa Postal 68528, CEP 21941-972, Rio de Janeiro, RJ, Brazil}



\begin{abstract}
The production of axionlike particles (ALPs) with small masses in ultraperipheral $Pb - p$ and $Pb - Pb$ collisions at the LHC is investigated. The cross section and {kinematical distributions } associated to the diphoton final state produced in the $\gamma \gamma \rightarrow a \rightarrow \gamma \gamma$ subprocesses are estimated considering a realistic set of kinematical cuts. A detailed analysis of the backgrounds is performed and the expected sensitivity to the ALP production is derived. Our results demonstrate that a future experimental analysis of the exclusive  diphoton production for the forward rapidities probed by the LHCb detector can improve the existing exclusion limits on the ALP - photon coupling in the mass range 2 GeV $\le m_a \le$ 5 GeV. 
\end{abstract}


\pacs{}

\keywords{Axiolike particles, Light-by-light scattering, Photon -- Photon interactions, Heavy ion collisions}

\maketitle

\begin{figure}
\begin{tabular}{ccc}
\hspace{-0.6cm}
{\psfig{figure=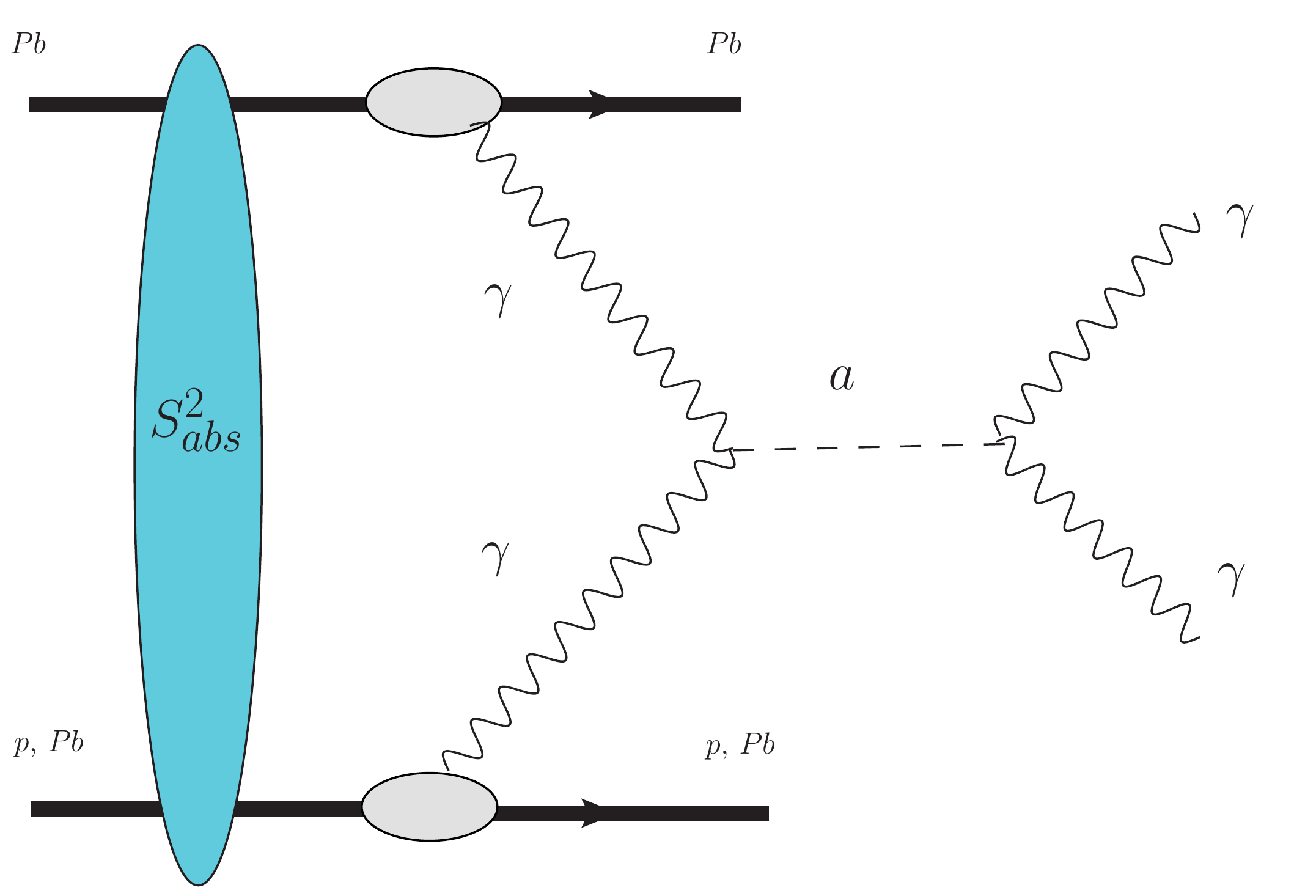,width=7.0cm}} &
{\psfig{figure=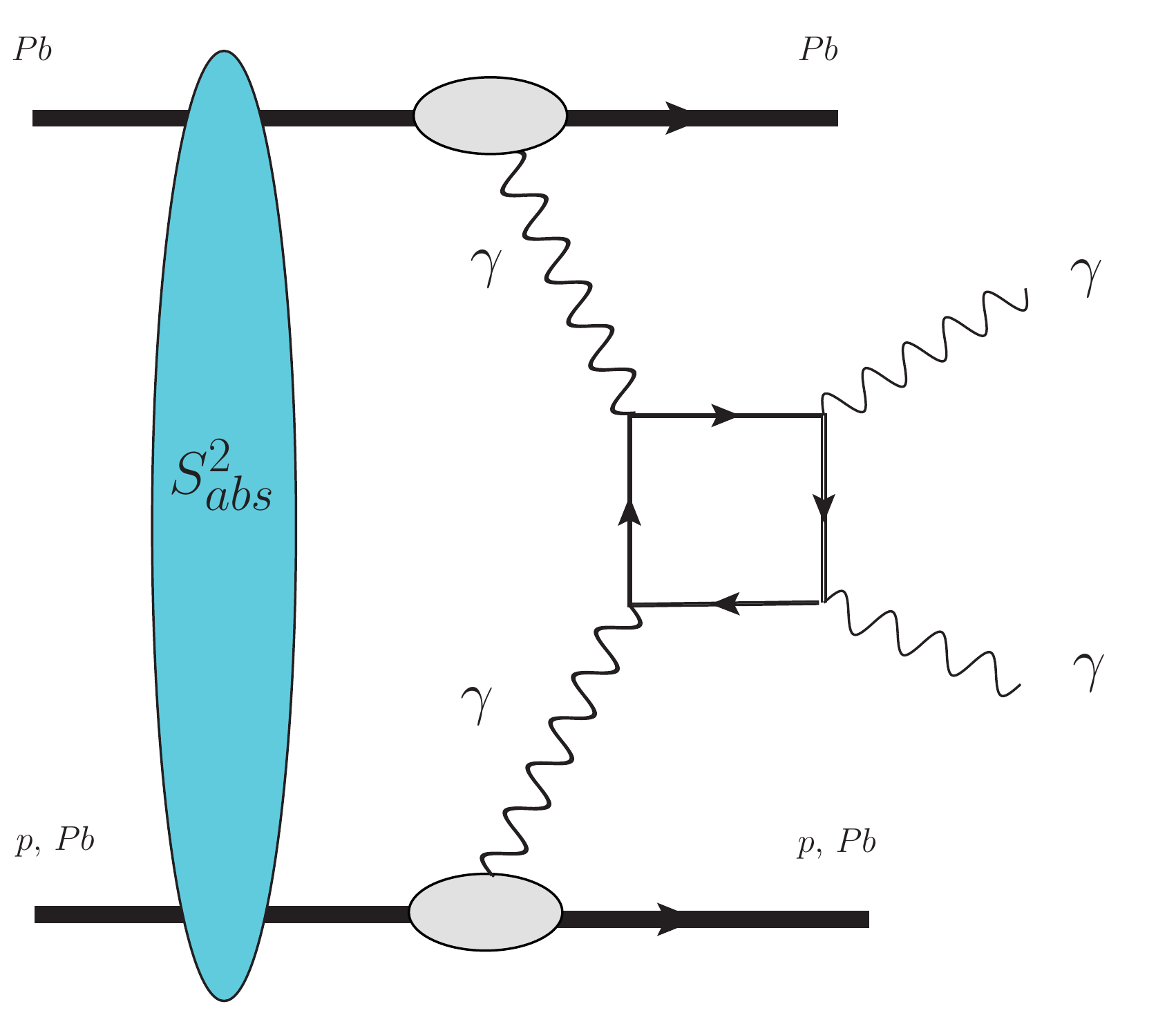,width=5.5cm}} &
{\psfig{figure=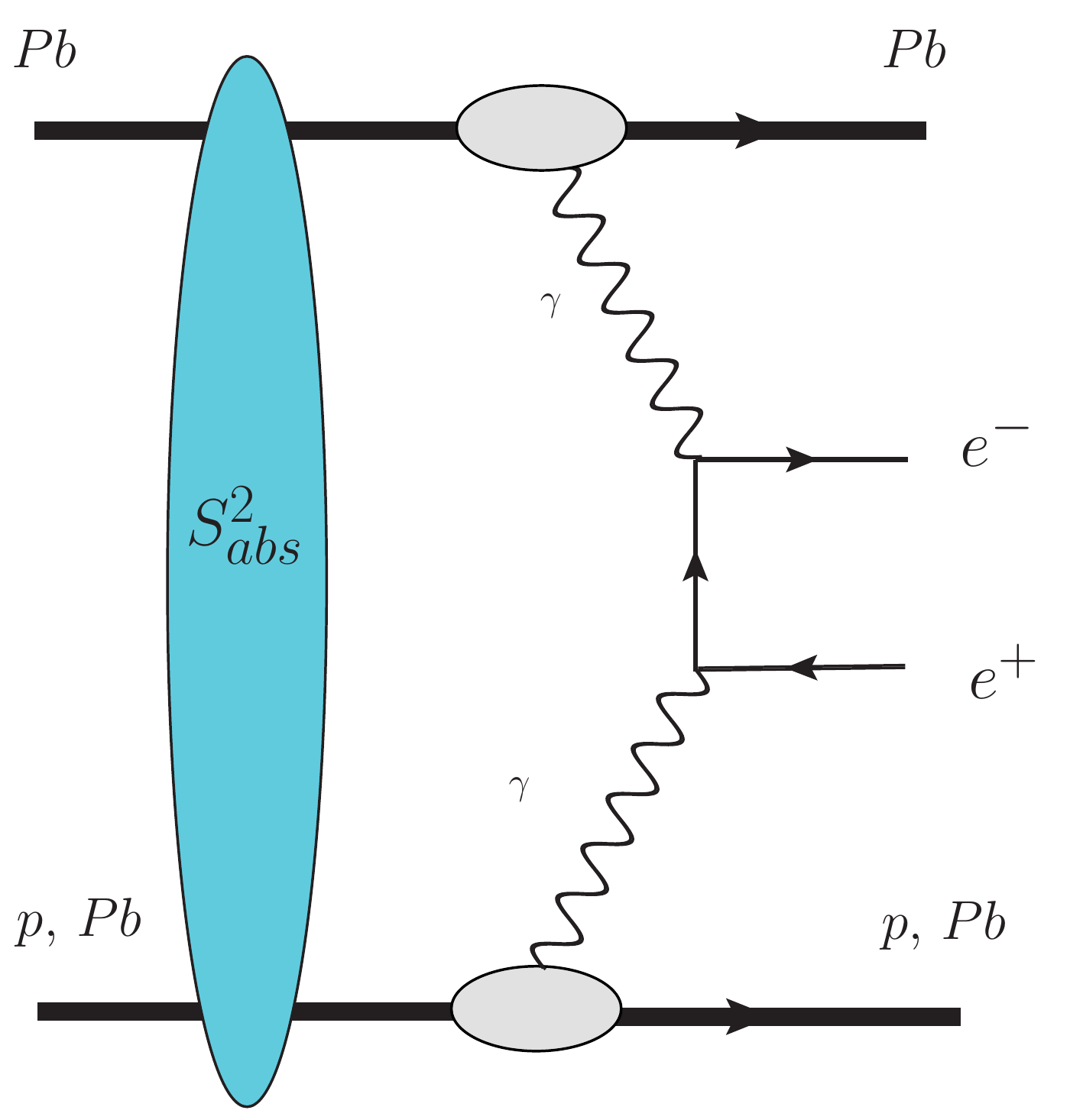,width=4.5cm}}  \\
(a) & (b) & (c)  
\end{tabular}                                                                                                                       
\caption{Diphoton production in $Pb - p/Pb - Pb$ collisions by (a) the $\gamma \gamma \rightarrow a \rightarrow\gamma \gamma$ subprocess and (b) Light -- by -- Light scattering. The dielectron production, which is a important background when the electrons are misidentified as photons, is represented in the diagram (c).}
\label{fig:diagram}
\end{figure}

The possibility of search for new physics  in ultraperipheral heavy ion collisions (UPCs) has been largely  discussed  in recent years (For recent reviews see, e.g. Refs. \cite{klein,Bruce:2018yzs}). Such studies are strongly motivated by the large photon - photon luminosity ($\propto Z_1^2 Z_2^2$, where $Z_i$ are the atomic numbers of the incident particles) and the clean environment present in UPCs \cite{upc}, which enhance the production cross sections and reduce the  backgrounds, allowing to derive stringent limits on some Beyond Standard Model (BSM) scenarios that predict  new particles that couple predominantly to photons. In particular, the recent results reported by the ATLAS and CMS Collaborations for the diphoton production in UPCs \cite{atlas_alp,cms_alp}  provide the strongest limits to date in the properties  of the axionlike particles (ALPs) in the mass region 5 GeV $\le m_a \le$ 100 GeV. Such pseudo - scalar particles are pseudo -- Nambu -- Goldstone bosons predicted to occur in many extensions of the Standard Model (SM), as e.g. supersymmetry or string theories, which can be produced at colliders and decay into photons, charged leptons, light hadrons or jets, depending on the ALP mass and coupling structure (See e.g. Ref. \cite{review_axion}).  The coupling of  the pseudoscalar ALP $a$ to photons is usually described by a Lagrangian of the form
\begin{equation}
\mathcal{L}=\frac{1}{2}\partial^\mu a \partial_\mu a -\frac{1}{2}m_a^2 a^2 -\frac{1}{4}g_{a\gamma} a F^{\mu\nu}\tilde{F}_{\mu\nu}\;,
\end{equation}
where $m_a$ is the ALP mass, $g_{a\gamma}$ is the coupling constant to photons and  $\tilde{F}^{\mu\nu}  = \frac{1}{2}\epsilon^{\mu\nu \alpha\beta} F_{\alpha\beta}$, which implies that ALPs can be produced by the photon -- photon fusion and can decay into a diphoton system, as represented in Fig. \ref{fig:diagram} (a), where the incident particles are assumed as source of photons. The presence of an axionlike particle as a resonance in the $\gamma \gamma \rightarrow a \rightarrow \gamma \gamma$ subprocess is expect to modify the scattering rates of the Light - by - Light (LbL) scattering, characterized by the $\gamma \gamma  \rightarrow \gamma \gamma$ subprocess and represented in Fig. \ref{fig:diagram} (b), with a larger impact on the invariant mass distribution of the photon pair system. Such aspect was exploited in the studies performed in Refs. \cite{knapen,royon,nos_plb} as well as in the experimental analysis at the LHC \cite{atlas_alp,cms_alp}. In particular, in the exploratory study performed in Ref. \cite{nos_plb}, we have demonstrated that a forward detector, as the LHCb, is ideal to probe 	an ALP with small mass ($m_a  \lesssim 10$ GeV), which is a mass region of difficult experimental access for the ATLAS and CMS detectors due to restrictions on the photon reconstructive capabilities. In Ref. \cite{nos_plb} we have restricted our analysis for $Pb - Pb$ collisions and studied the ALP production assuming four different combinations for the ALP mass and coupling. 
 One of the goals of this letter is to extend the analysis for ultraperipheral $Pb - p$ collisions at $\sqrtsnn = 8.1$ TeV. Moreover, we complement the previous study for $Pb - Pb$ collisions by presenting the results for a larger number of combinations of ALP mass and coupling, with a special focus on the low mass region $m_a  \lesssim 5$ GeV and forward rapidities, which can be reached by the LHCb detector. For completeness, we also will present the predictions considering that the diphoton system is produced at central rapidities and that the low invariant masses range could be reached by a central detector in the forthcoming years. Another goal is to derive, for the first time, the exclusion limits on the ALP parameter space ($m_a, \, g_{a\gamma}$) that can be established by studying the production of diphotons with small invariant masses in ultraperipheral $Pb - p$ and $Pb - Pb$ collisions at the LHC. In our analysis, we will assume that the main backgrounds for the ALP production come from the LbL scattering [Fig. \ref{fig:diagram} (b)] and the dielectron production [Fig. \ref{fig:diagram} (b)], where both the electron and positron are misidentified as photons. As demonstrated in Refs. \cite{nos_plb,nos_diphoton,vicwer}, the background associated to the diphoton production in gluon - induced processes is negligible when the exclusivity cuts are taken into account. The background associated to photoproduction of pseudoscalar resonances and  $\pi^0$ - pairs was addressed in Ref. \cite{mariola} and its contribution  can, in principle, be strongly reduced through the inclusion of additional cuts on the transverse momenta of the two photons.  In our study, the predictions for the signal and backgrounds will be derived using the \superchic 4 Monte Carlo event generator \cite{superchic4} including 
realistic cuts on rapidity, invariant mass, transverse momentum and acoplanarity of the diphoton system.

In the equivalent photon approximation \cite{epa}, the total cross section for the hadronic process 
can be factorized in terms of the equivalent photon spectrum  
of the incident nuclei and the elementary cross section for the $\gamma \gamma \rightarrow a \rightarrow\gamma \gamma$ process as follows
\begin{eqnarray}
\sigma \left(h_1 h_2 \rightarrow h_1 \otimes \gamma \gamma \otimes h_2;\sqrtsnn \right)   
&=& \int \mbox{d}^{2} {\mathbf r_{1}}
\mbox{d}^{2} {\mathbf r_{2}} 
\mbox{d}W 
\mbox{d}y \frac{W}{2} \, \hat{\sigma}\left(\gamma \gamma \rightarrow a \rightarrow \gamma \gamma ; 
W \right )  N_{h_1}\left(\omega_{1},{\mathbf r_{1}}  \right )
 N_{h_2}\left(\omega_{2},{\mathbf r_{2}}  \right ) S^2_{abs}({\mathbf b})  
  \,\,, \nonumber \\
\label{cross-sec-2}
\end{eqnarray}
where $h_i = p$ or $Pb$, $\sqrtsnn$ is center - of - mass energy of the $h_1h_2$ collision, $\otimes$ characterizes a rapidity gap in the final state, 
$W = \sqrt{4 \omega_1 \omega_2} = m_{\gamma \gamma}$ is the invariant mass of the $\gamma \gamma$ system  and $y$ its rapidity. Moreover,
$N_{h_i}(\omega_i, {\mathbf r}_i)$ is the equivalent photon spectrum associated to hadron $h_i$, which allows to estimate the number the photons  with energy $\omega_i$ at a transverse distance ${\mathbf r}_i$  from the center of hadron, defined in the plane transverse to the trajectory, which is determined by the charge form factor of the particle. The absorptive factor $S^2_{abs}({\mathbf b})$, which depends on the impact parameter ${\mathbf b}$ of the $h_1 h_2$ collision, is included in order to exclude the overlap between the colliding hadrons and insure the dominance of the electromagnetic interaction. Following Ref.\cite{superchic3}, we will assume that the photon spectrum can be expressed in terms of the electric form factor  and that the absorptive corrections $S^2_{abs}({\mathbf b})$ for $\gamma \gamma$ interactions in $Pb - p$ and $Pb - Pb$ collisions  can be estimated taking into account the multiple scatterings between the nucleons of the incident hadrons, which allows to calculate the probability for no additional  hadron -- hadron rescattering at different impact parameters. As in Refs. \cite{knapen,nos_plb}, the cross section for the $\gamma \gamma \rightarrow a \rightarrow \gamma \gamma$ subprocess will be estimated assuming that the ALP is a narrow resonance with a mass $m_a$ that couples to the $\gamma \gamma$ system with strength $g_{a\gamma}$. Similar expressions can be written for the LbL scattering and dilepton production, with the difference that they will be expressed in terms of 
$\hat{\sigma}\left(\gamma \gamma  \rightarrow \gamma \gamma \right )$ and $\hat{\sigma}\left(\gamma \gamma  \rightarrow e^+ e^- \right )$, respectively (For details see, e.g. Refs. \cite{nos_diphoton,nos_dilepton}). The signal and the background associated to the LbL scattering will be characterized by a very clean final state, consisting  of the diphoton system,  two intact hadrons and  two rapidity gaps, i.e. empty regions  in pseudo-rapidity that separate the intact very forward hadrons from the $\gamma \gamma$ system. A similar final state will be observed in the dilepton production if the electron and positron are both misidentified as photons. Although the probability of this misidentification is expected to be small ($\approx 0.5\%$ for each individual lepton), such background can become nonnegligible due to the huge cross section for the dielectron production in UPCs (See e.g. Ref. \cite{nos_dilepton}).  
As in Refs. \cite{nos_plb,nos_diphoton,nos_dilepton}, we will estimate the total cross sections and differential distributions using the 
\superchic  MC event generator \cite{superchic4,superchic3}. The LbL scattering will be calculated taking into account of the fermion loop contributions as well as  the contribution from $W$ bosons, with the later being negligible for the values of $m_{\gamma \gamma}$ of interest in our analysis. We will focus on the diphoton production with invariant mass in the range $1 \le m_{\gamma \gamma} \le 20$ GeV, which will allow us to probe the production of ALPs with small mass. In particular, we will present results for eight distinct combinations of mass and coupling, with special emphasis on the mass range $m_a \le 5$ GeV, which cannot currently be  reached by the ATLAS and CMS detectors. We will select events in which $m_{\gamma \gamma}$ > 1 GeV and $p_{T}(\gamma,\gamma)$ > 0.2 GeV, where $p_T$ is the transverse momentum of the photons. Moreover, we will impose a cut on  the acoplanarity ($1 -(\Delta \phi/\pi)$ < 0.01) and  transverse momentum of the diphoton system  ($p_{T}(\gamma \gamma)$ < 0.1 GeV). As demonstrated in Refs. \cite{nos_plb,nos_diphoton,nos_dilepton}), such cuts are able to suppress the background associated with gluon - initiated processes and to guarantee the exclusivity criteria. As we are interested in the exclusive ALP production in the kinematical range probed by the LHCb detector, we  will select events where photons are produced  in the rapidity range $2.0 < |\eta(\gamma^{1},\gamma^{2})| < 4.5$ with 0 extra tracks with $p_{T} > 0.1$ GeV in the  range $5.5 < |\eta| <  8.0$, which  corresponds to the HERSCHEL acceptance installed in the LHCb experiment~\cite{herschel}. {For completeness of our analysis, we also will present predictions for a typical central detector, assuming that it is able to observe events of a diphoton system with a small invariant mass.} For this case, 
we only will select events where photons are produced in the rapidity range $|\eta(\gamma^{1},\gamma^{2})| < 2.5$ with 0 extra tracks.

\begin{center}
\begin{table}
\scalebox{0.8}{
\begin{tabular}{|c|c|c|c|c|c|c|c|c|c|c|}
\hline 
{\bf $Pb-Pb$ @ $\sqrtsnn$ = 5.5 TeV} & {\bf LbL} & {\bf $ e^{+}e^{-} (\gamma \gamma) $} &\multicolumn{8}{c|}{{\bf ALP}}   \tabularnewline
\hline 
\, &  \,  & \, &\multicolumn{2}{c|}{  $m_{a}=2$ GeV} & \multicolumn{2}{c|}{  $m_{a}=3$ GeV}& \multicolumn{2}{c|}{ $m_{a}=4$ GeV} & \multicolumn{2}{c|}{ $m_{a}=5$ GeV}  \tabularnewline
\hline 
Coupling ($g_{a\gamma}$)[GeV$^{-1}$] & -  & - & $1\cdot10^{-3}$  &$8\cdot10^{-4}$ & $1\cdot10^{-3}$  &$8\cdot 10^{-4}$&$1\cdot10^{-3}$&$8\cdot10^{-4}$&$1\cdot10^{-3}$&$8\cdot10^{-4}$\tabularnewline 
\hline 
\hline
{\bf Generation level} & \multicolumn{10}{c|}{\,} \tabularnewline
\hline 
Total Cross section {[}nb{]} & 18000.0  & 13000.0 & 17640.0 & 11288.0 &13000.0 &8369.0  & 11000.0  & 6914.0 &  8944.0  & 5725.0     \tabularnewline
\hline 
\hline
{\bf Exclusivity cuts} & \multicolumn{10}{c|}{\,} \tabularnewline
\hline
$m_{\gamma\gamma}> 1\:\rm{GeV}, p_{T}(\gamma,\gamma)>0.2\:\rm{GeV}$& 13559.0  & 2500.0  & 17245.0  &11035.0 & 12873.0  & 8289.0& 10928.0  & 6869.0  &  8916.0 & 5707.0     \tabularnewline
\hline 
$1- (\Delta \phi/\pi) < 0.01$ &  8834.0 & 1550.0  & 13217.3  &  8458.0 & 11033.0  & 7102.0 & 9846.0  &  6189.0  & 8389.5    &  5370.1   \tabularnewline
\hline 
$p_{T}(\gamma\gamma)< 0.1$  GeV & 8826.0 & 1550.0  & 13206.0  & 8450.5  & 11019.0  & 7092.0 & 9827.0  & 6177.0  & 8369.0   & 5357.0     \tabularnewline
\hline 
\hline
{\bf Forward selection} & \multicolumn{10}{c|}{\,} \tabularnewline
\hline
 $2.0<\eta(\gamma,\gamma)<4.5$             & 616.0  &87.5   & 1282.2& 820.5 & 974.0 & 614.0  &784.0 & 493.0& 580.3  & 371.4     \tabularnewline
\hline
\hline
$1.5 < m\left(\gamma\gamma\right) < 2.5  $ & 166.0 & 23.5  & 1282.2 & 820.5  & - &- &-&-& -  & -    \tabularnewline
\hline
$2.5 < m\left(\gamma\gamma\right) < 3.5$   & 33.0  &  5.0 &  -& -  & 974.0 & 614.0   & -&- &  - & -   \tabularnewline
\hline 
$3.5 < m\left(\gamma\gamma\right) < 4.5 $  & 11.3  &  1.8 &  -& - & - &- & 784.0 & 493.0  & -  & -   \tabularnewline
\hline 
$4.5 < m\left(\gamma\gamma\right) < 5.5  $ & 5.9   & 0.8  & - & - & -  &-&- &- & 580.3 & 371.4 \tabularnewline
\hline
\hline 
{\bf Central selection} & \multicolumn{10}{c|}{\,} \tabularnewline
\hline
 $|\eta(\gamma,\gamma)|<2.5$ & 4730.0  & 750.0 & 8053.5  & 5154.0 & 6990.0 & 4500.0 & 6422.0 & 4036.0 & 5633.4  &  3606.0    \tabularnewline
\hline
\hline
$1.5 < m\left(\gamma\gamma\right) < 2.5  $ &1361.6&  222.5 & 8053.5   & 5154.0  & - &- &-&-& -  & -    \tabularnewline
\hline
$2.5 < m\left(\gamma\gamma\right) < 3.5$ & 314.0  &  52.5 &  -& - & 6990.0  &4500.0 & -&- &  - & -   \tabularnewline
\hline 
$3.5 < m\left(\gamma\gamma\right) < 4.5 $ &125.1  & 20.0  &  -& - & - &- & 6422.0 & 4036.0 & -  & -   \tabularnewline
\hline 
$4.5 < m\left(\gamma\gamma\right) < 5.5  $ &79.0  & 10.0  & - & - & -  &-&- &- & 5633.4 & 3606.0 \tabularnewline
\hline
\end{tabular}}
\caption{Predictions for the cross sections for the exclusive diphoton production in $Pb - Pb$ collisions at $\sqrtsnn = 5.5$ TeV associated to the  ALP production with different masses $m_a$ and couplings $g_{a\gamma}$, LbL  scattering  and exclusive dielectron production adjusted by the misidentification probability for each electron individually. }
\label{tab:PbPb}
\end{table}
\end{center}

\begin{center}
\begin{table}
\scalebox{0.8}{
\begin{tabular}{|c|c|c|c|c|c|c|c|c|c|c|}
\hline 
{\bf $Pb-p$ @ $\sqrtsnn$ = 8.1 TeV} & {\bf LbL} & {\bf $ e^{+}e^{-} (\gamma \gamma) $} &\multicolumn{8}{c|}{{\bf ALP}}   \tabularnewline
\hline 
\, &  \,  & \, &\multicolumn{2}{c|}{  $m_{a}=2$ GeV} & \multicolumn{2}{c|}{  $m_{a}=3$ GeV}& \multicolumn{2}{c|}{ $m_{a}=4$ GeV} & \multicolumn{2}{c|}{ $m_{a}=5$ GeV}  \tabularnewline
\hline 
Coupling ($g_{a\gamma}$)[GeV$^{-1}$] & -  & - & $1\cdot10^{-3}$  &$8\cdot10^{-4}$ & $1\cdot10^{-3}$  &$8\cdot 10^{-4}$&$1\cdot10^{-3}$&$8\cdot10^{-4}$&$1\cdot10^{-3}$&$8\cdot10^{-4}$\tabularnewline 
\hline 
\hline
{\bf Generation level} & \multicolumn{10}{c|}{\,} \tabularnewline
\hline 
Total Cross section {[}pb{]} & 5913.0 & $4250.0$&6053.0 &3874.0 &4894.0 & 3132.0 & 4170.0  & 2669.0 & 3661.1   & 2343.0    \tabularnewline
\hline 
\hline
{\bf Exclusivity cuts} & \multicolumn{10}{c|}{\,} \tabularnewline
\hline
$m_{\gamma\gamma}> 1\:\rm{GeV}, p_{T}(\gamma,\gamma)>0.2\:\rm{GeV}$&4392.2  &$825.0$  & 5890.0  &3769.5 &4837.3  &3096.0 & 4143.0 & 2652.0 &3646.7   & 2333.8    \tabularnewline
\hline 
$1- (\Delta \phi/\pi) < 0.01$ & 1926.5 & $350.0$ &2974.0   &  1903.4 &2737.2 & 1752.0 & 2535.0 & 1622.2& 2362.0   & 1511.7    \tabularnewline
\hline 
$p_{T}(\gamma\gamma)< 0.1$  GeV & 1845.5 & $325.0$  & 2832.0& 1812.3&2554.0 & 1634.0& 2312.3 & 1480.0& 2104.0   & 1347.0     \tabularnewline
\hline 
\hline
{\bf Forward selection} & \multicolumn{10}{c|}{\,} \tabularnewline
\hline
 $2.0<\eta(\gamma,\gamma)<4.5$ &189.0  &24.3 & 396.0  &253.5 &365.1 &  234.0&336.0 & 215.0& 309.3  & 197.9     \tabularnewline
\hline
\hline
$1.5 < m\left(\gamma\gamma\right) < 2.5  $ &53.2 & 7.0 &396.0& 253.5  & - &- &-&-& -  & -    \tabularnewline
\hline
$2.5 < m\left(\gamma\gamma\right) < 3.5$ & 11.4&$1.6$ &  -& - & 365.1 & 234.0& -&- &  - & -   \tabularnewline
\hline 
$3.5 < m\left(\gamma\gamma\right) < 4.5 $ &5.6 &$0.8$  &  -& - & - &- & 336.0&  215.0& -  & -   \tabularnewline
\hline 
$4.5 < m\left(\gamma\gamma\right) < 5.5  $ & 3.0& 0.5  & - & - & -  &-&- &- & 309.3 & 197.9 \tabularnewline
\hline
\hline
{\bf Central selection} & \multicolumn{10}{c|}{\,} \tabularnewline
\hline
 $|\eta(\gamma,\gamma)|<2.5$ &872.0  &142.5 & 1548.0  &991.1 &1421.4 &  910.0&1309.0 & 838.0& 1209.0  & 774.0     \tabularnewline
\hline
\hline
$1.5 < m\left(\gamma\gamma\right) < 2.5  $ &253.1 & 42.5 &1548.0& 991.1  & - &- &-&-& -  & -    \tabularnewline
\hline
$2.5 < m\left(\gamma\gamma\right) < 3.5$ & 58.1&12.3 &  -& - & 1421.4 & 910.0& -&- &  - & -   \tabularnewline
\hline 
$3.5 < m\left(\gamma\gamma\right) < 4.5 $ &27.2 &4.0  &  -& - & - &- & 1309.0& 838.0& -  & -   \tabularnewline
\hline 
$4.5 < m\left(\gamma\gamma\right) < 5.5  $ & 14.9& 2.2  & - & - & -  &-&- &- & 1209.0 & 774.0 \tabularnewline
\hline
\end{tabular}}
\caption{Predictions for the cross sections for the exclusive diphoton production in $Pb - p$ collisions at $\sqrtsnn = 8.1$ TeV associated to the  ALP production with different masses $m_a$ and couplings $g_{a\gamma}$, LbL  scattering  and exclusive dielectron production adjusted by the misidentification probability for each electron individually. }
\label{tab:pPb}
\end{table}
\end{center}

In Table \ref{tab:PbPb} we present our predictions for the total cross sections associated to the signal and backgrounds considered in our analysis  assuming $Pb - Pb$ collisions at $\sqrtsnn = 5.5$ TeV. The background that comes from photon misidentification due to electrons is denoted by $e^+ e^- (\gamma \gamma)$, which is estimated assuming that the associated  probability is $0.5\%$ for each individual lepton. We consider four distinct values for the ALP mass $m_a$ and two possible values for the coupling constant $g_{a\gamma}$, which complement the predictions presented in Ref. \cite{nos_plb}. The results at the generation level are also presented in order to illustrate the impact of the exclusivity cuts. Our results indicate that the background associated to $e^+ e^- (\gamma \gamma)$ is strongly suppressed by these cuts. For $m_a = 2$ GeV, the signal dominates for both values of $g_{a\gamma}$, with the dominance being larger  when an additional cut on the invariant mass is applied. For larger values of the ALP mass, the predictions for the signal and the LbL background are similar, but again the selection of events in a given $m_{\gamma \gamma}$ range has a large impact, strongly suppressing the LbL background. Similar conclusions are valid for $Pb - p$ collisions at $\sqrtsnn = 8.1$ TeV, as observed from the analysis of the results presented in Table \ref{tab:pPb}. The main difference is associated to the fact that the  cross sections for $Pb - p$ collisions are of the order of pb, while for $Pb - Pb$ they are ${\cal{O}}$(nb). Such result is directly associated to the dependence of the photon - photon luminosity on the product $Z_1^2 Z^2_2$, which implies that the $Pb - p$ predictions are suppressed by a factor $\approx (82)^2$ in comparison to those for $Pb - Pb$. However, it is important to emphasize that the expected integrated luminosities for the future $Pb - p$ runs are, in general, larger than for $Pb - Pb$ collisions, which partially reduces this suppression in the associated number of events and become $Pb - p$ collisions competitive to probe the ALP production.

\begin{center}
\begin{figure}[t]
\includegraphics[width=0.3\textwidth]{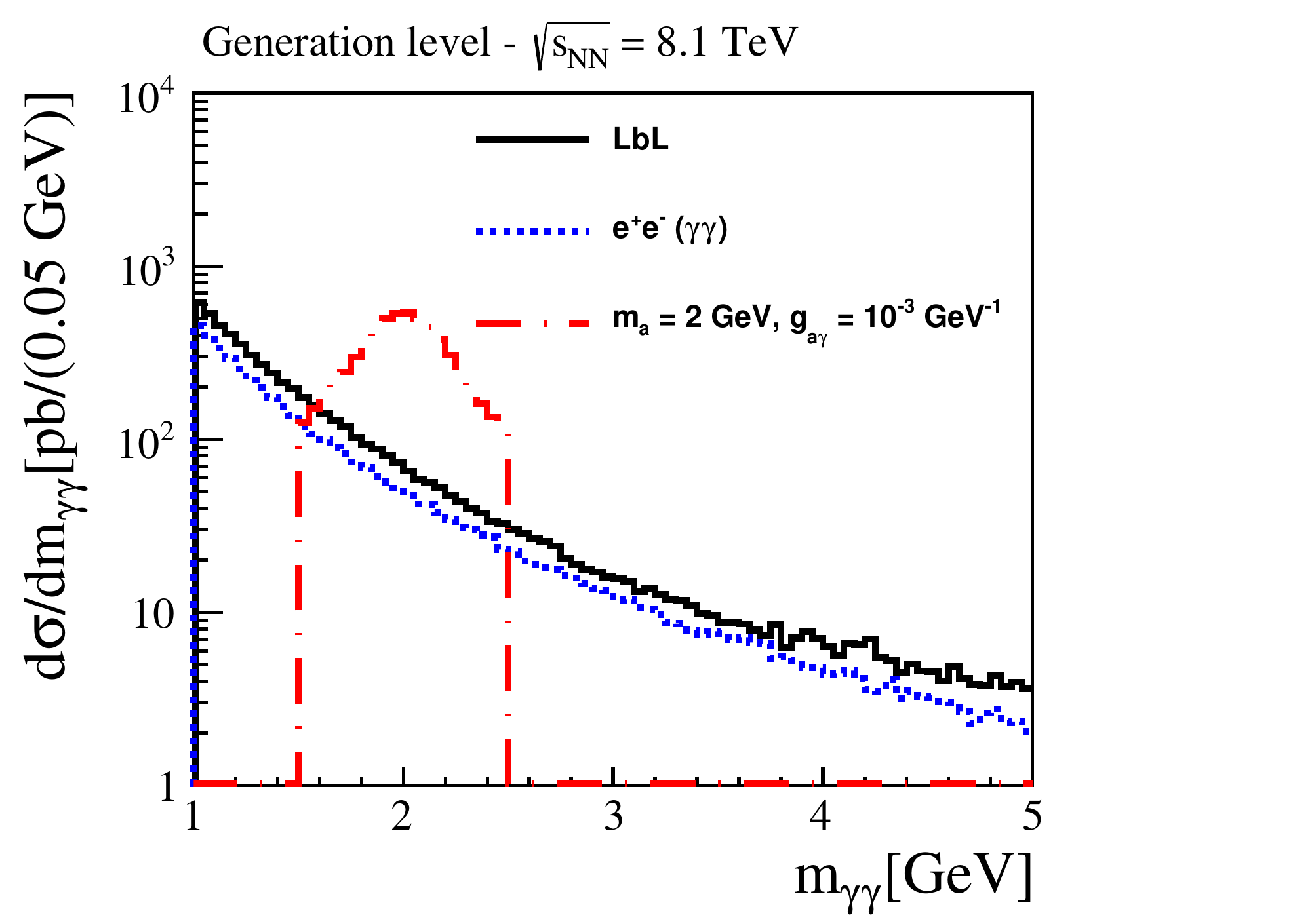}
\includegraphics[width=0.3\textwidth]{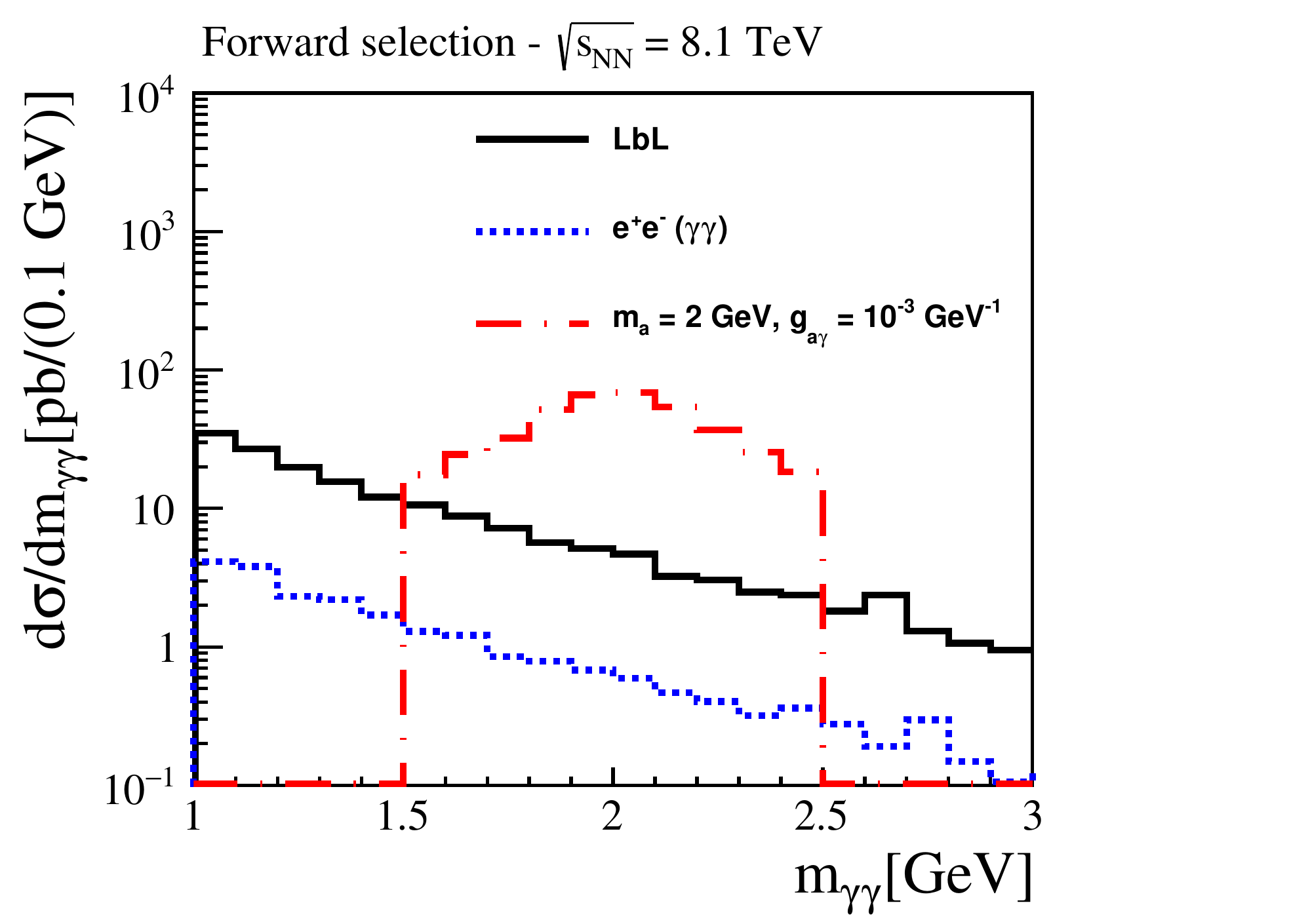}
\includegraphics[width=0.3\textwidth]{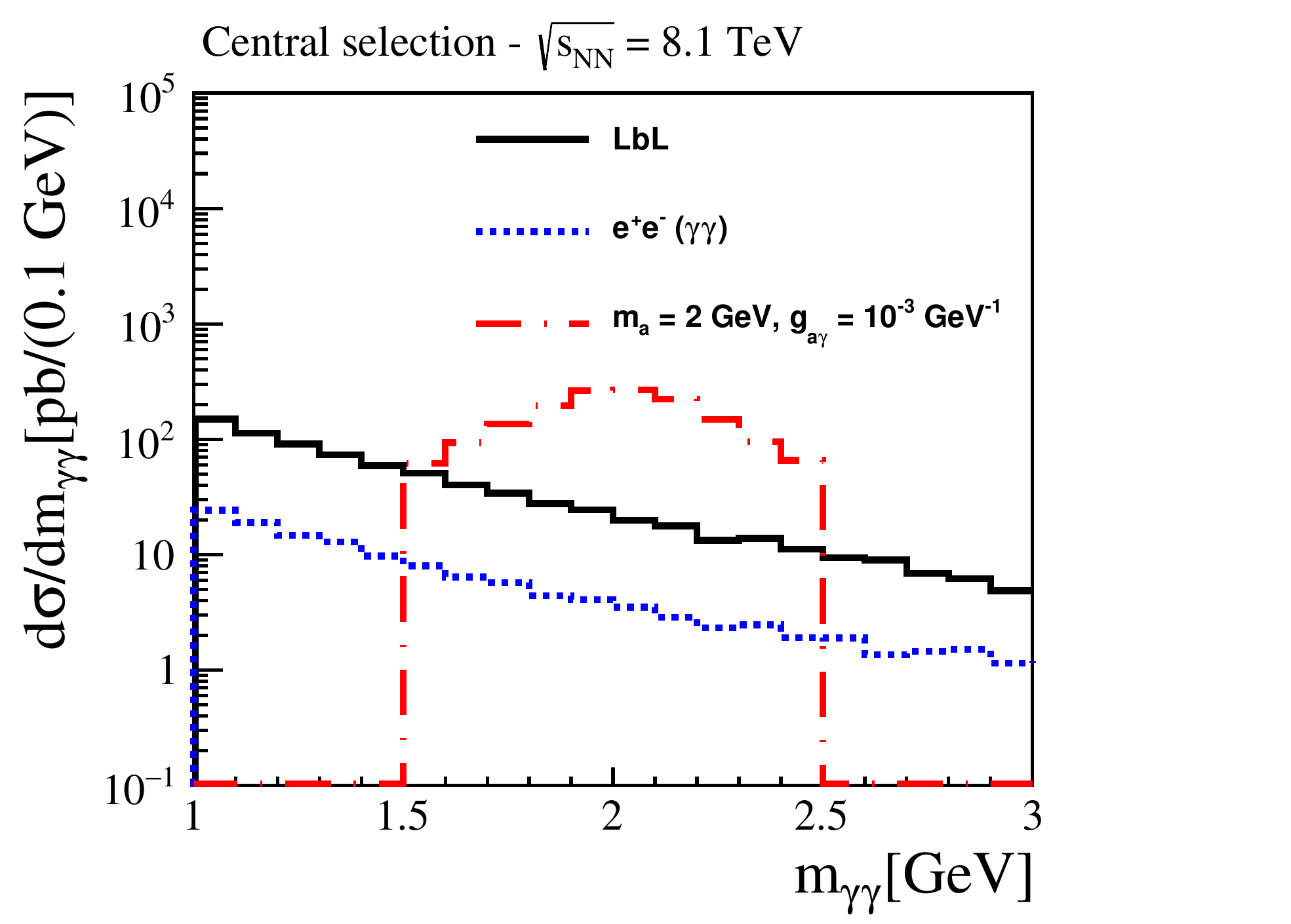}
\includegraphics[width=0.3\textwidth]{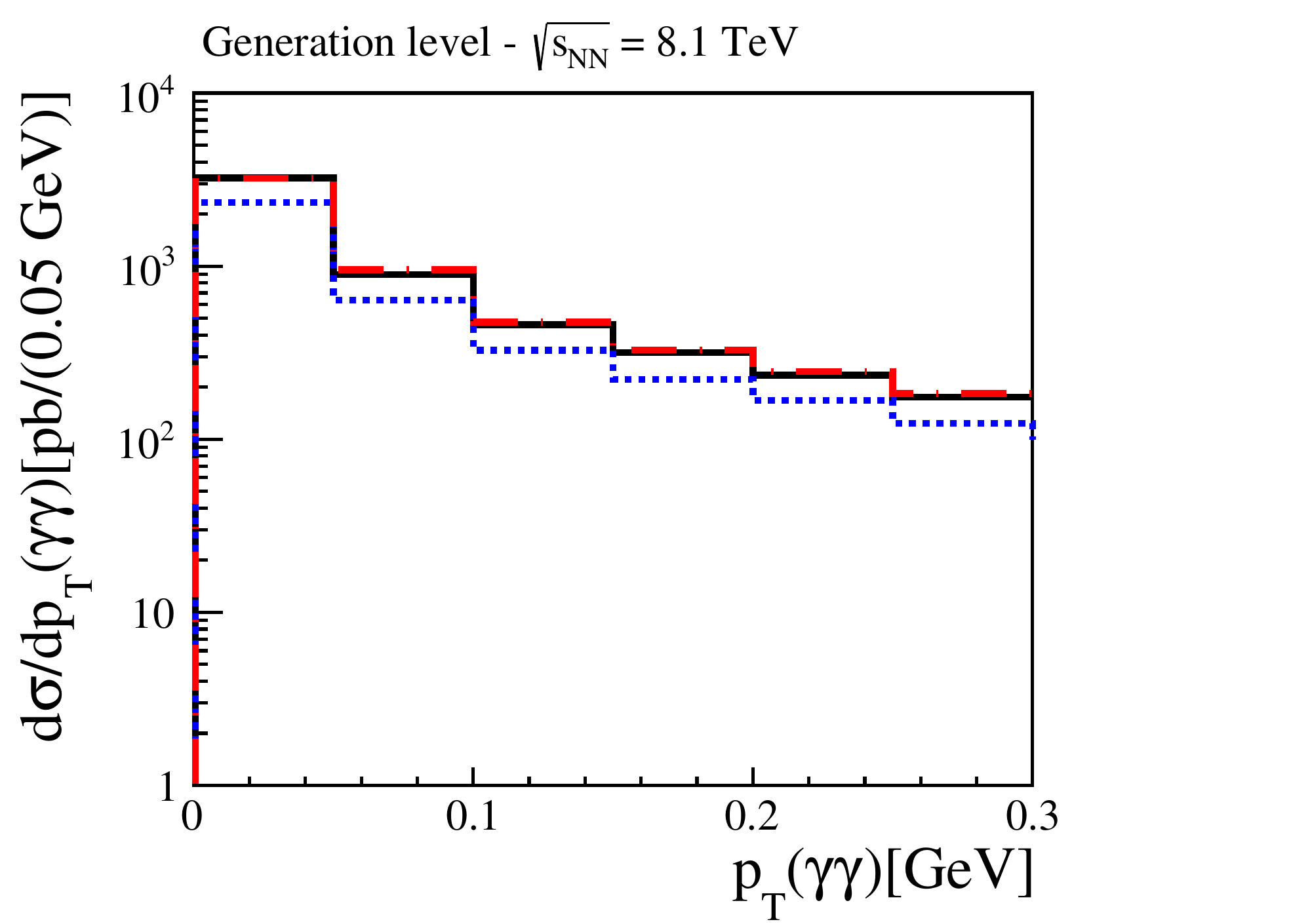}
\includegraphics[width=0.3\textwidth]{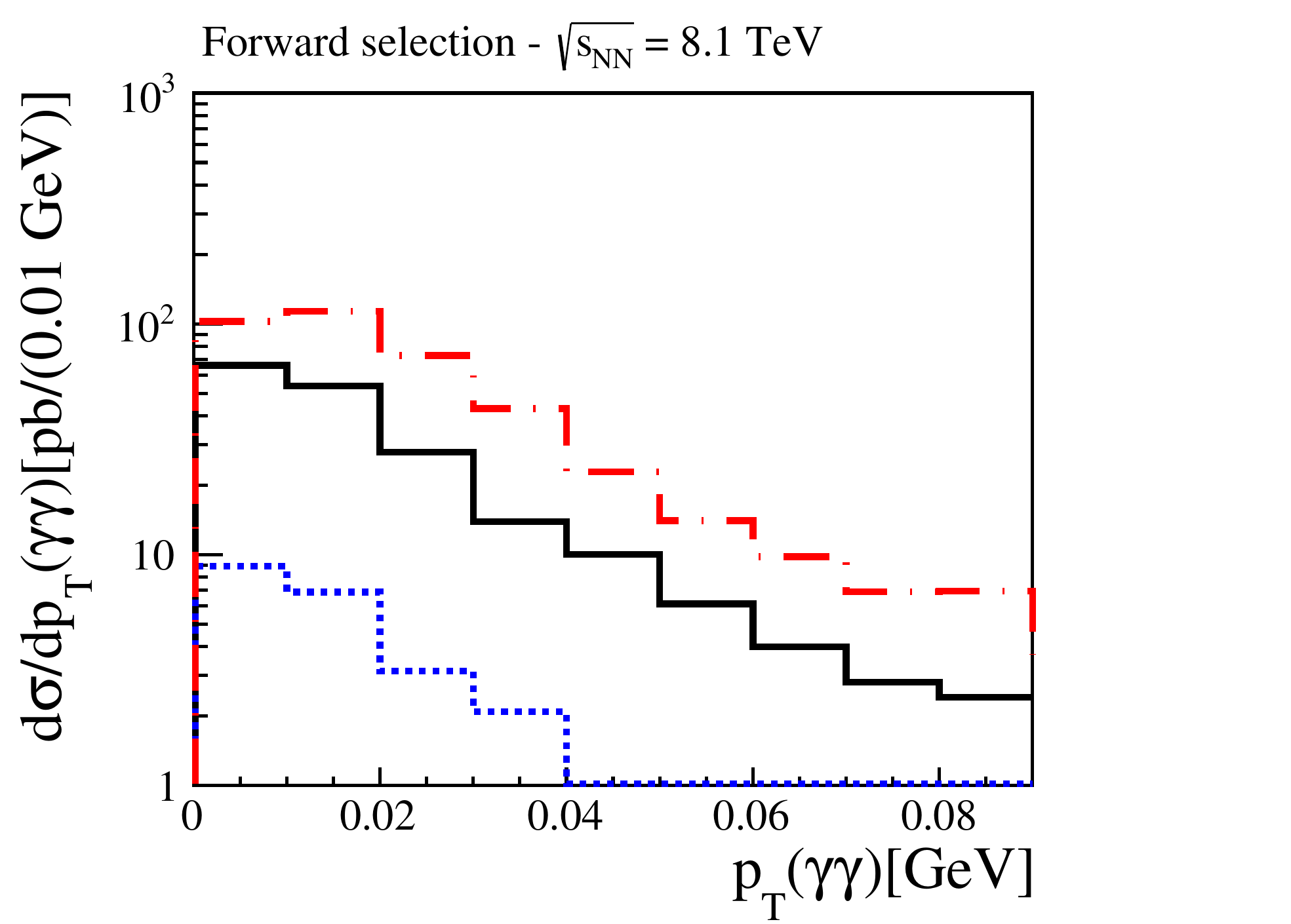}
\includegraphics[width=0.3\textwidth]{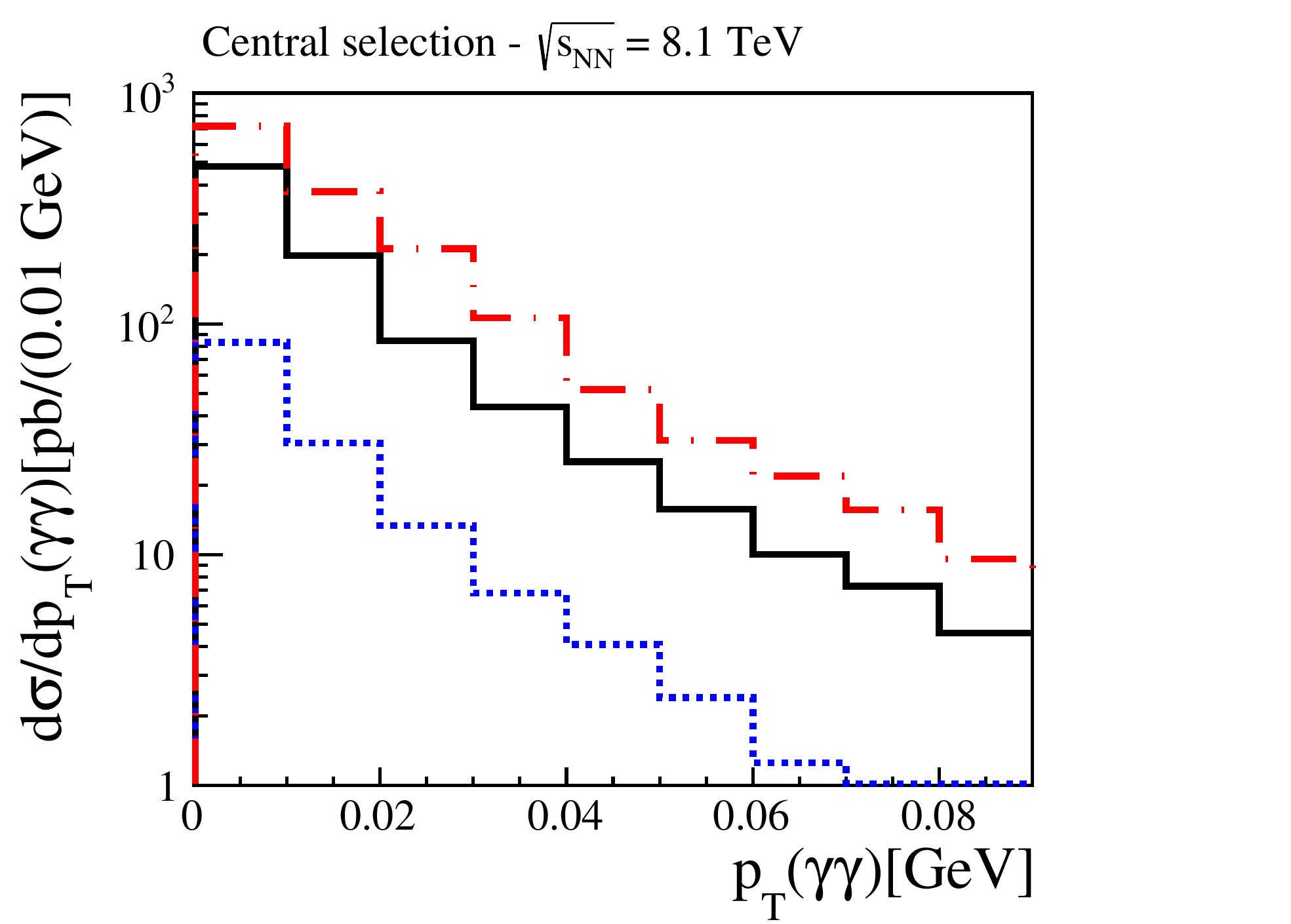}
\includegraphics[width=0.3\textwidth]{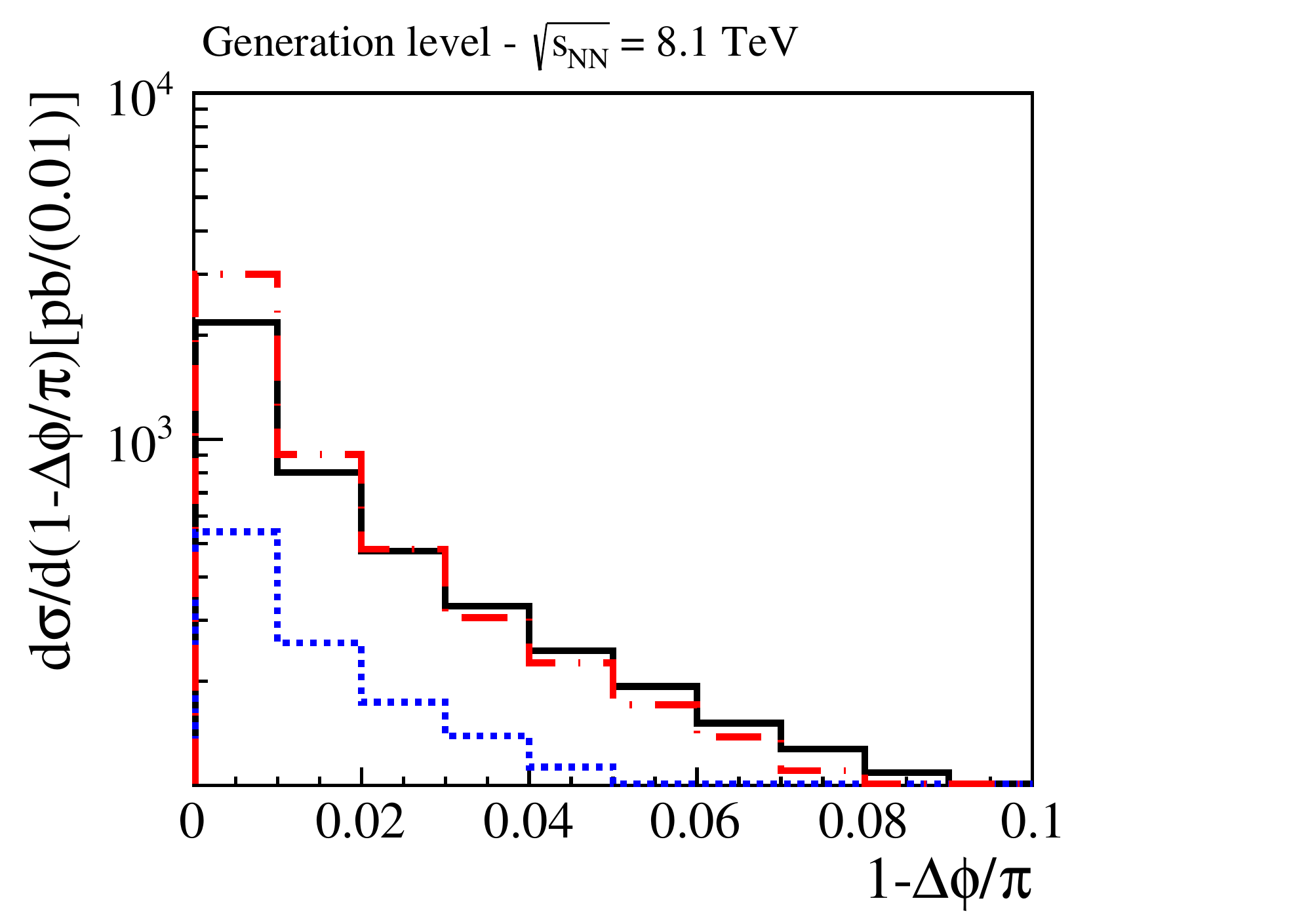}
\includegraphics[width=0.3\textwidth]{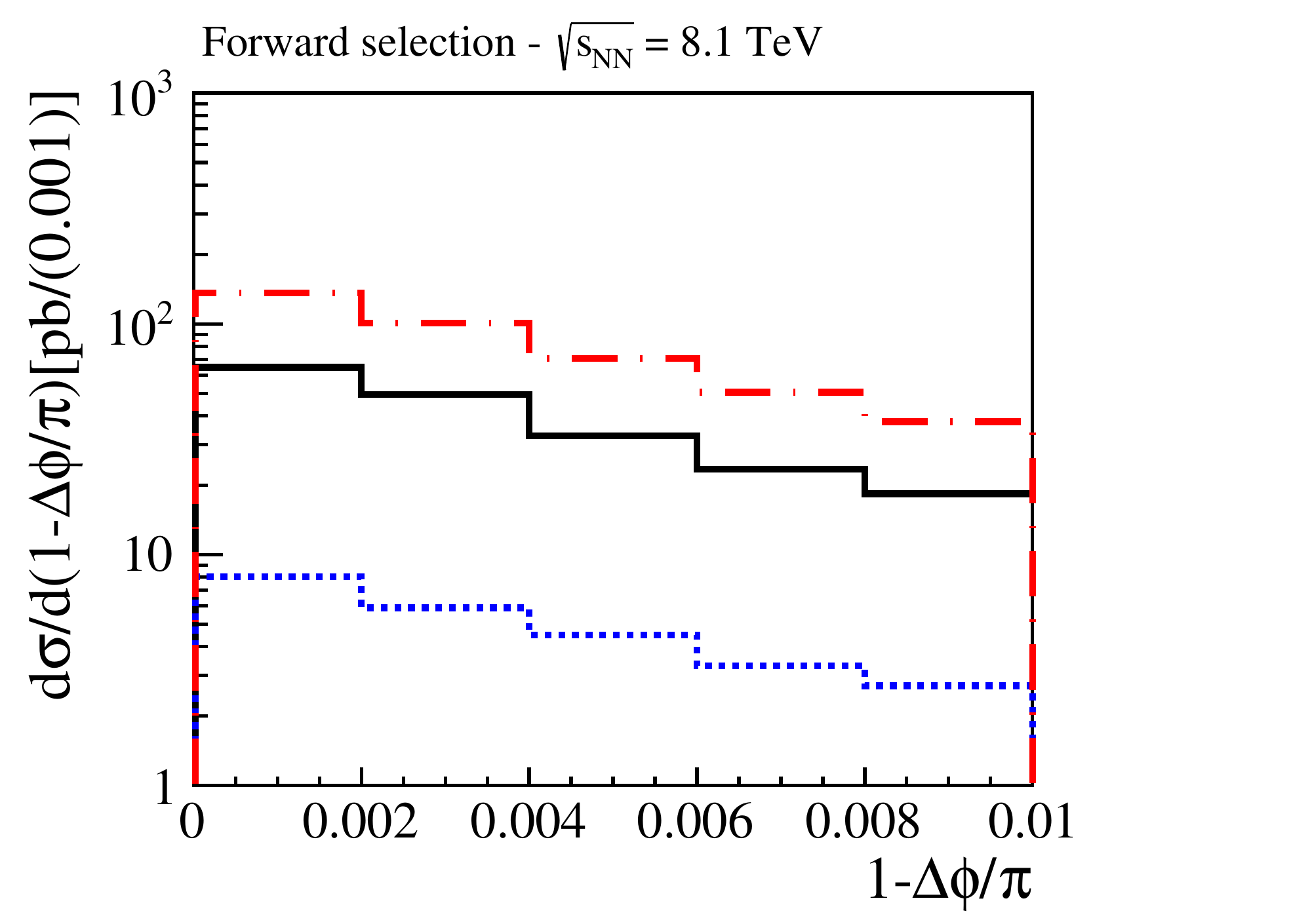}
\includegraphics[width=0.3\textwidth]{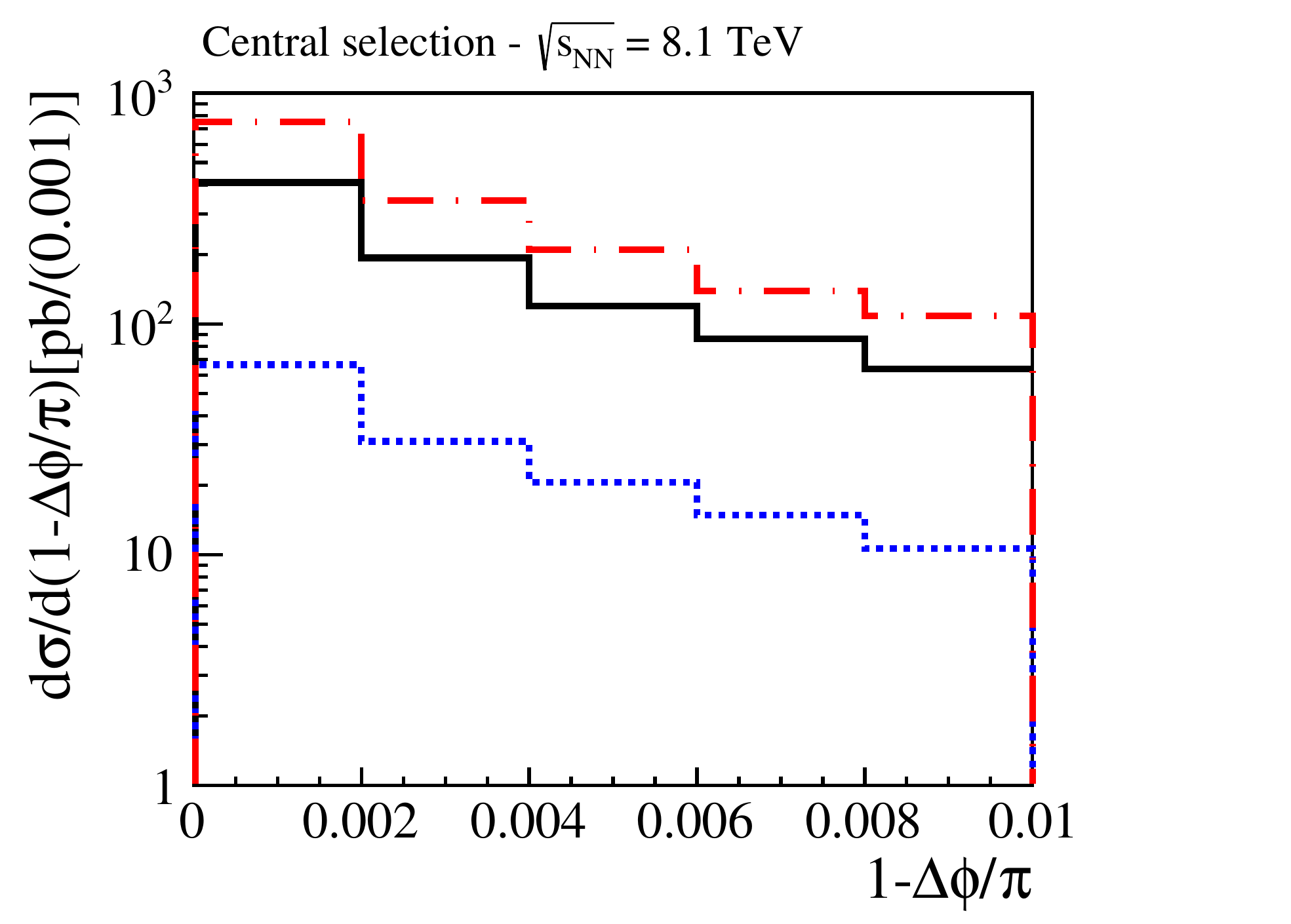}
\includegraphics[width=0.3\textwidth]{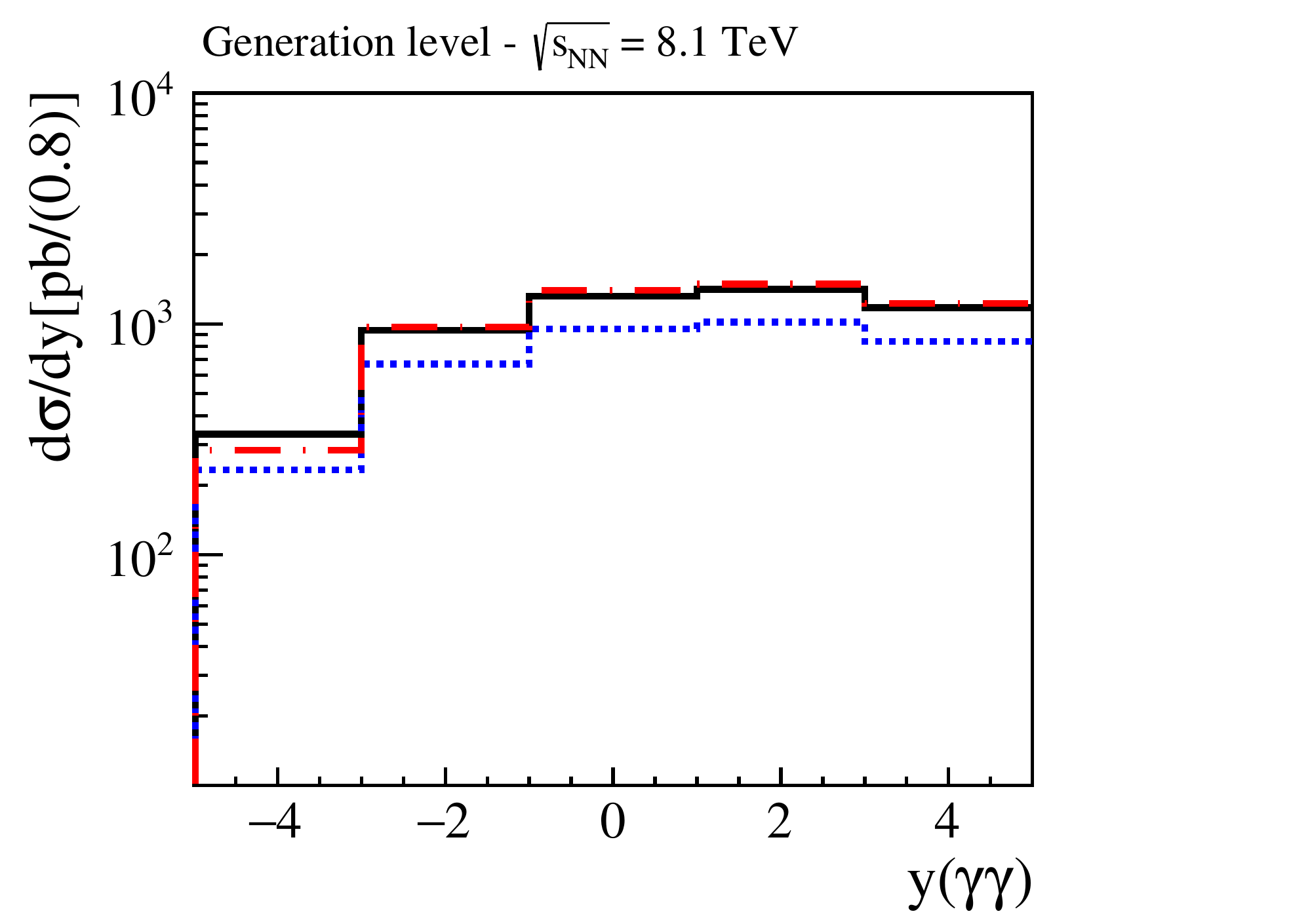}
\includegraphics[width=0.3\textwidth]{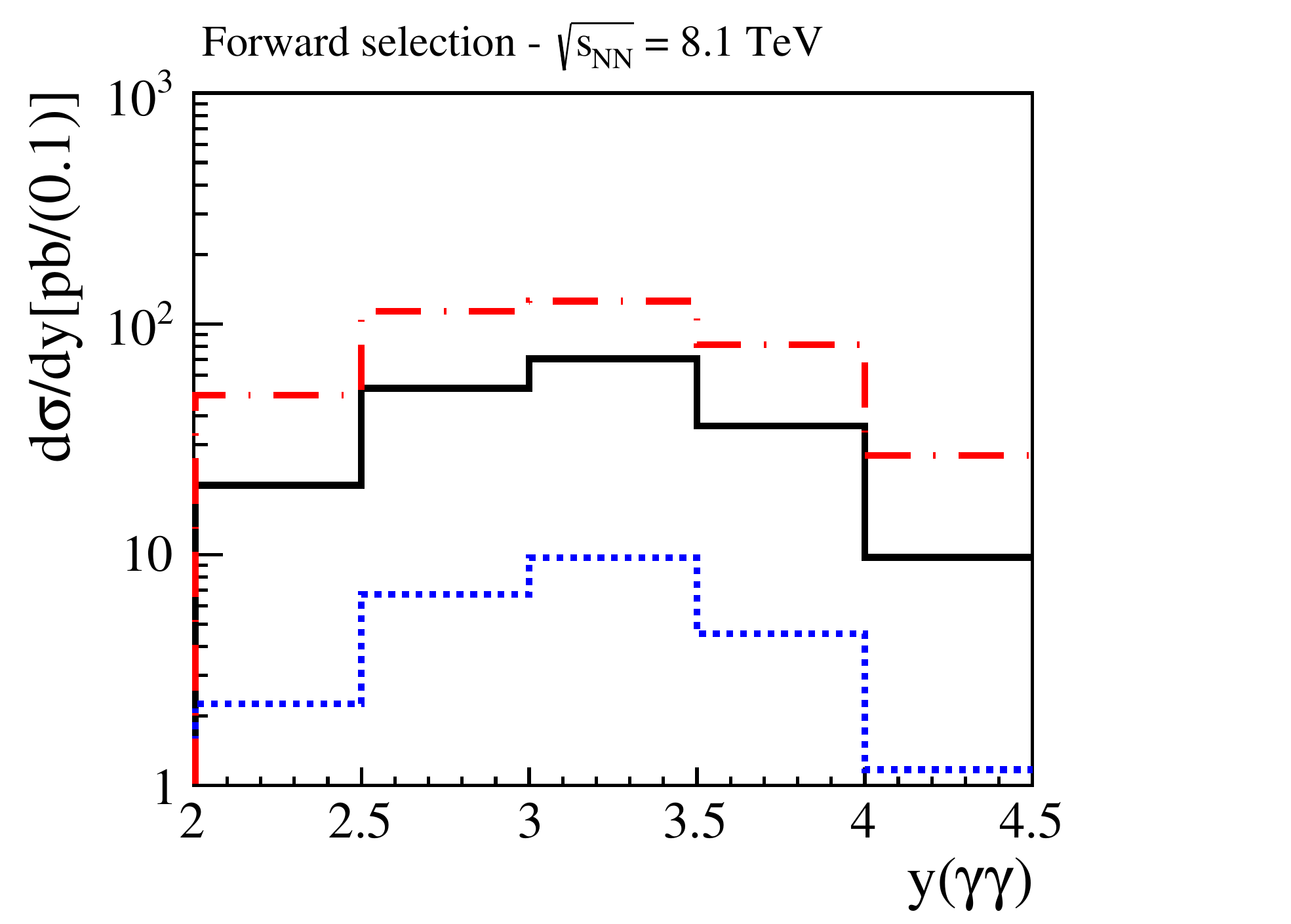}
\includegraphics[width=0.3\textwidth]{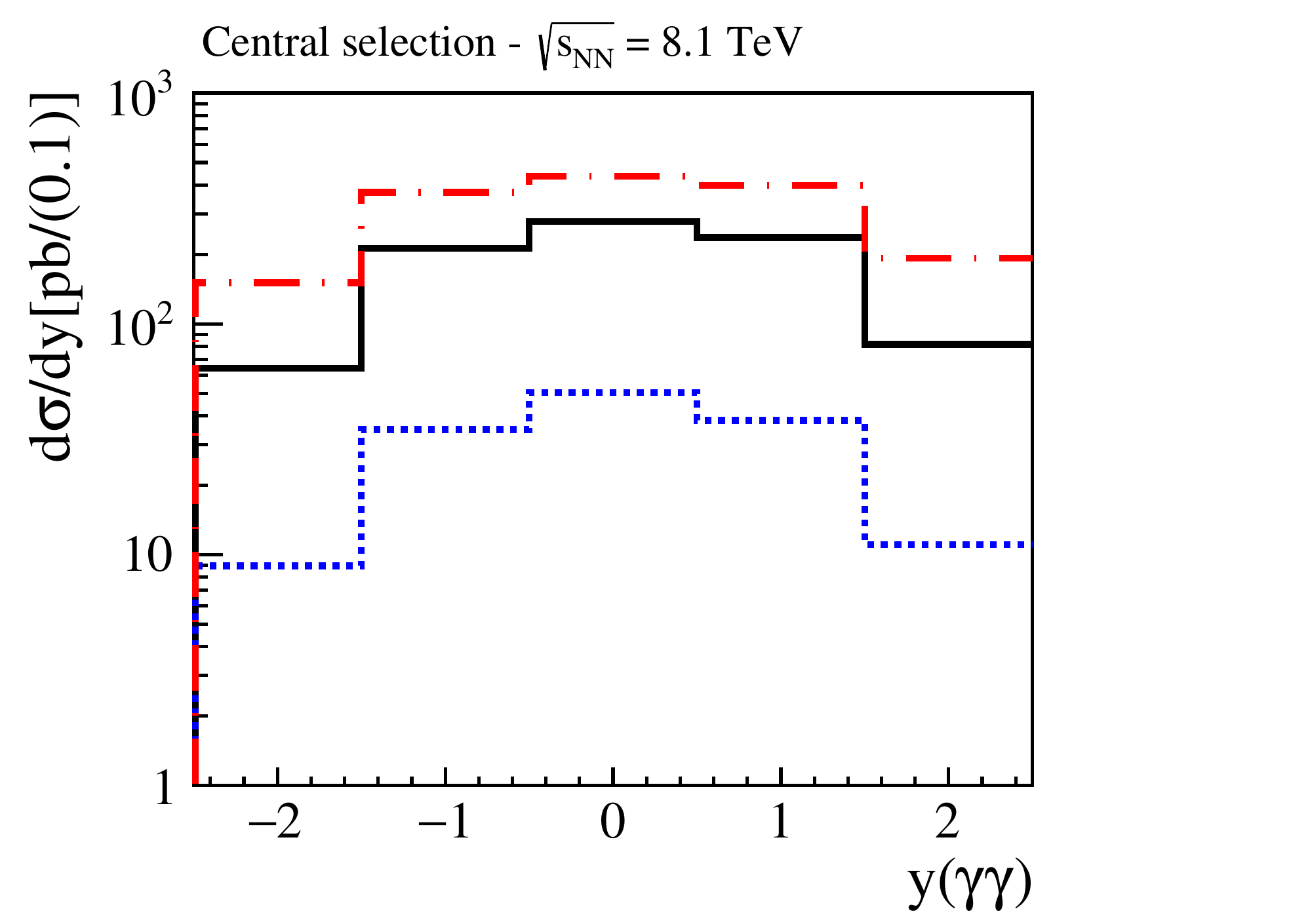}
\caption{Predictions for the  invariant mass $m_{\gamma\gamma}$, transverse momentum $p_{T}(\gamma\gamma)$, acoplanarity and rapidity distributions of the diphoton system produced in  $Pb - p$ collisions at $\sqrtsnn = 8.1$ TeV derived  at generation level (left panels) and considering the forward (middle panels) and central (right panels) selections.}
\label{fig:dist}
\end{figure}
\end{center}

The predictions presented in Tables \ref{tab:PbPb} and \ref{tab:pPb} are calculated using realistic experimental requirements and they indicate that the searching for ALPs with small masses in the exclusive diphoton production is, in principle, feasible by the LHCb collaboration. Motivated by this result, in Fig. 
\ref{fig:dist} (middle panels) we present our predictions for the invariant mass, transverse momentum, acoplanarity and rapidity distributions considering $Pb - p$ collisions  at $\sqrtsnn = 8.1$ TeV and the forward selection. For completeness, we also present the predictions at the generation level (left panels) and for the central selection (right panels). The results were derived for $m_a = 2$ GeV and $g_{a\gamma} = 10^{-3}$ GeV$^{-1}$. Similar distributions for $Pb - Pb$ collisions can be found in Ref. \cite{nos_plb}. One has that 
the ALP production as a resonance in the $s$ - channel of the $\gamma \gamma \rightarrow \gamma \gamma$ reaction implies a peak in the invariant mass distribution.  In agreement with the results presented in Table \ref{tab:pPb}, the contribution of the $e^+ e^- (\gamma \gamma)$ background is of the same order of the signal at the generation level, but is strongly suppressed by the exclusivity cuts. Similar suppression is observed for the LbL background. Moreover, the signal dominantes the invariant mass distribution if the range 1.5 GeV $ \le m_{\gamma \gamma} \le 2.5$ GeV is selected, which implies that the study of the diphoton production at small masses is a direct probe of the ALP. The shape of the transverse momentum, acoplanarity and rapidity distributions for the signal and backgrounds are similar, differing in the normalization.

\begin{center}
\begin{figure}[t]
\includegraphics[width=0.45\textwidth]{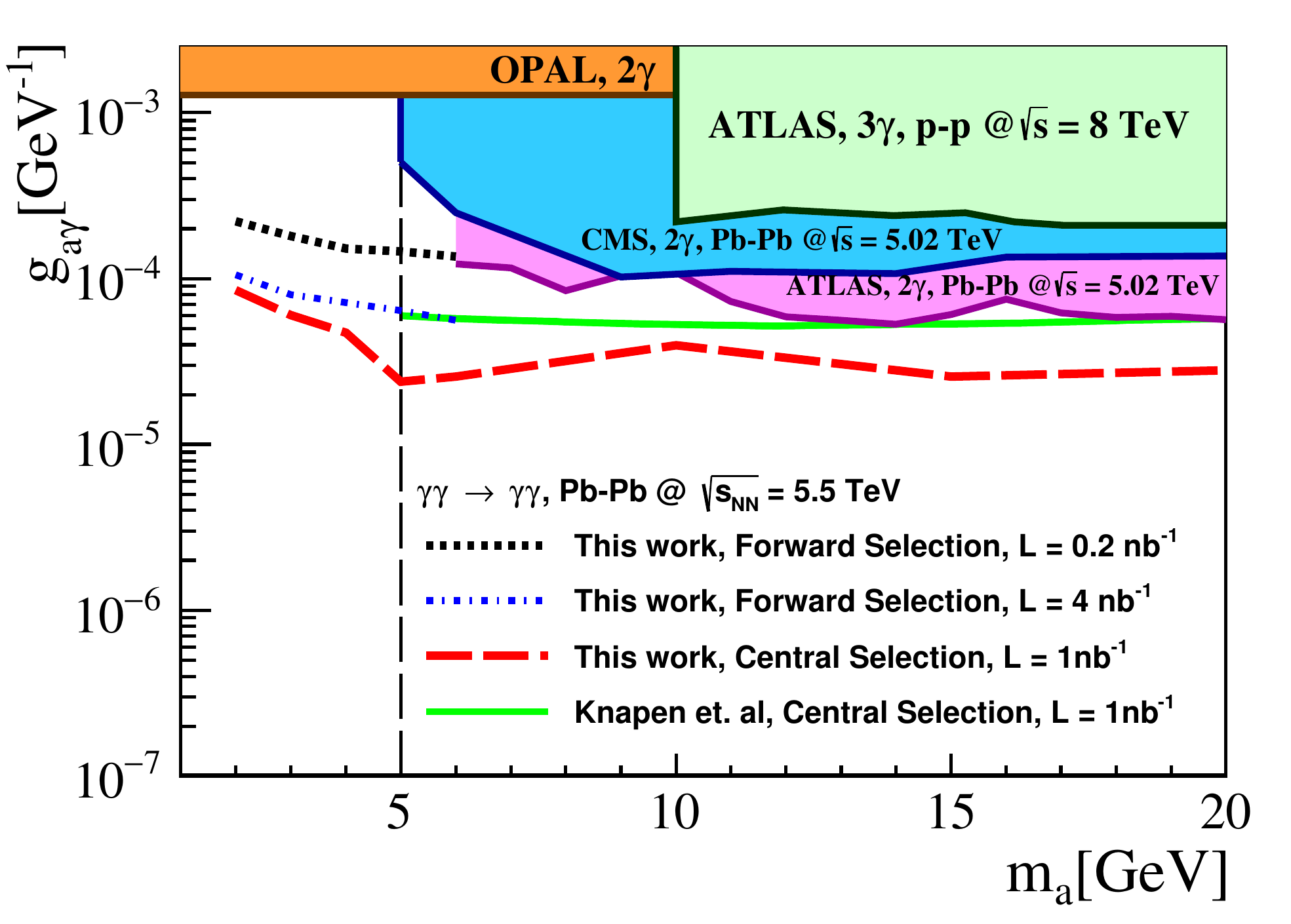}
\includegraphics[width=0.45\textwidth]{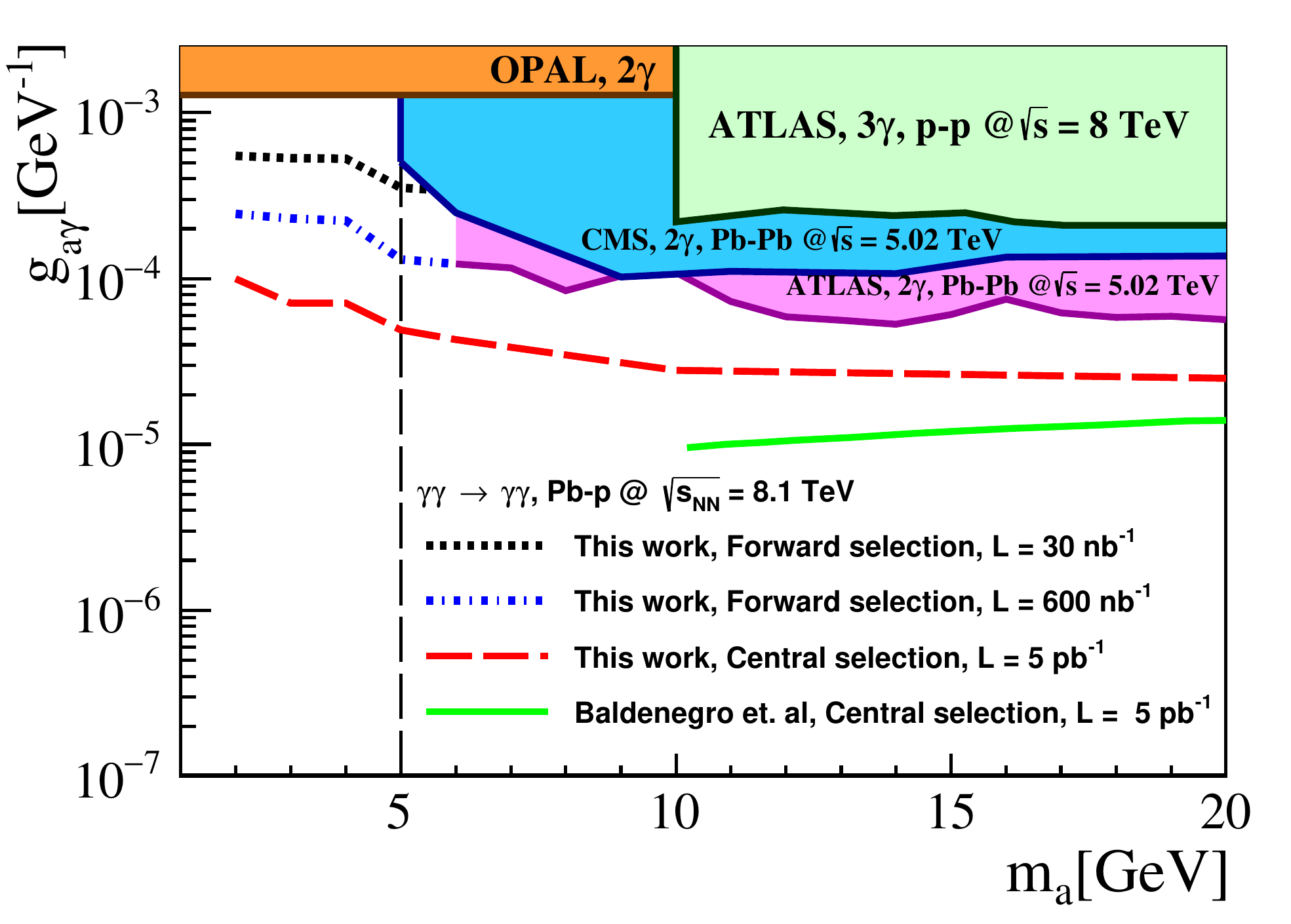}
\caption{Expected sensitivity to the ALP production in ultraperipheral $Pb - Pb$ (left panel) and $Pb - p$ (right panel) collisions at the LHC derived considering different luminosities and selection criteria. Existing exclusion limits derived in $e^+ e^-$, $pp$ and $PbPb$ collisions are represented by the various shaded regions. }
\label{fig:limits}
\end{figure}
\end{center}

Finally, let's estimate the expected sensitivity to ALPs in the exclusive diphoton production in ultraperipheral $Pb - p$ and $Pb - Pb$ collisions considering the current and expected future scenarios for the integrated luminosity. The existing exclusion limits on the ALP parameter space ($m_a, \, g_{a\gamma}$) for ALP  masses in the range of interest for this study has been derived assuming that ${\cal{B}}(a \rightarrow \gamma \gamma) = 1$ and come from $e^+ e^-$ collisions at the LEP \cite{opal2gamma} and hadronic collisions at the LHC \cite{atlas_alp,cms_alp,atlas3gamma}. They are represented by the shaded regions in Fig. \ref{fig:limits}. 
{In order to derive our projections, we set a 95$\%$ confidence limit bound on the signal cross section and remaining backgrounds, assuming Poisson statistics as 
implemented in previous studies~\cite{knapen,royon}.
} 
Our results for $Pb - Pb$ collisions  at $\sqrtsnn = 5.5$ TeV and $Pb - p$ collisions at $\sqrtsnn = 8.1$ TeV are represented by dashed lines in the left and right panels of the Fig. \ref{fig:limits}, respectively. For comparison, we also present the projections derived in Refs. \cite{knapen,royon} for $m_a \geq 5$ GeV assuming the typical kinematical cuts that can be applied by the ATLAS and CMS detectors. Our results for the forward selection indicate that $Pb - Pb$ collisions provide more stringent constraints on the ALP - photon coupling than $Pb - p$ one for masses larger than 2 GeV. However, the analysis of the exclusive diphoton production in  $Pb - p$ collisions is also competitive to  exclude part of the ALP parameter space. These results indicate that a future analysis by the LHCb Collaboration is, in principle, able to enlarge  the exclusion limits in a region  not covered by previous experiments. For the central selection considered in our analysis, we derive projections slightly distinct from the results obtained in Refs. \cite{knapen,royon} for larger masses, which is associated to the different cuts implemented in the distinct studies. Our results point out that if a photon pair with low invariant masses could be measured by a central detector, we also shall improve the current bound limits in an unexploited region of the ALP parameter space.

In summary, we have performed  a  detailed analysis of the ALP production in the kinematical range probed by the LHCb detector and found that the study of the diphoton production in ultraperipheral $Pb - p$ and $Pb - Pb$ collisions can improve the limits on the ALP - photon couplings for ALP masses beyond the current reach of central detectors. Our results indicate that the exclusivity cuts suppress the potential backgrounds and that the ALP signal is dominant if the searching is performed for finite ranges of the invariant mass. Predictions for the rapidity, invariant mass, transverse momentum and acoplanarity distributions of the diphoton system produced in $Pb - p$ collisions were presented by the first time. Our study demonstrate that a future experimental analysis of the diphoton final state is a promissing observable to probe the existence of the Axionlike particles  and its properties.

\begin{acknowledgments}
This work was  partially financed by the Brazilian funding
agencies CNPq, CAPES,  FAPERGS, FAPERJ and INCT-FNA (process number 
464898/2014-5).
\end{acknowledgments}

 \end{document}